\documentclass[12pt]{iopart}

\usepackage{epsfig}
\begin{document}

\title{Edge magnetoplasmons in graphene }

\author{Ivana Petkovi\'{c}$^{1,2}\footnote{Present address: Department of Physics, Yale University, New Haven, CT 06520, USA }$, F.I.B. Williams$^{1,3}$, D. Christian Glattli$^1$}

\address{$^1$Service de Physique de l'\'{E}tat Condens\'{e}, Commissariat \`{a} l'\'{E}nergie Atomique, 91191 Gif-sur-Yvette, France\\$^2$Laboratoire National de M\'{e}trologie et d'Essais, 29 avenue Roger Hennequin, 78197 Trappes, France\\$^3$Institute for Solid State Physics and Optics, Wigner Research Centre for Physics, P.O. Box 49, H-1525 Budapest, Hungary}
\ead{ivana.petkovic@yale.edu}
\begin{abstract}
We have observed propagation of Edge Magneto-Plasmon (EMP) modes in graphene in
the Quantum Hall regime by performing 
picosecond time of
flight measurements between narrow contacts on the perimeter of micrometric exfoliated
graphene. We find the propagation to be chiral with low attenuation and to have a velocity which is quantized on
Hall plateaus. The velocity has two contributions, one arising from the Hall conductivity and
the other from carrier drift along the edge, which we were able to separate by their 
different filling factor dependence. The drift component is found to be slightly less than the Fermi velocity as expected for graphene dynamics in an abrupt edge potential. The Hall conduction
contribution is slower than expected and indicates a characteristic length in the Coulomb potential from the Hall charge of 
 about 500 nm. The experiment illustrates how EMP can be coupled to the electromagnetic field, opening the perspective of GHz to THz chiral plasmonics applications to devices 
such as voltage controlled phase shifters, circulators, switches and
compact, tunable ring resonators.
\end{abstract}

\maketitle

\section{Introduction}
Edge magnetoplasmons (EMP) are a set of elementary excitations of edge charge distribution which propagate along the periphery of 2-dimensional systems of charge in magnetic field. They arise by the action of the Lorentz force in accumulating charge against the constraining wall of the sample edge and can be thought of as a dynamic manifestation of the Hall effect, the symmetry properties of which are reflected in the chirality of propagation of the EMP. Their propagation velocity is a function of the Hall conductivity and of properties specific to the type of edge and particle dynamics. They have been observed in both classical\cite{Dahm, Glattli, Vinen} and quantum\cite{Stormer,Talyanskii,Grodnensky, Andrei, Wassermeier, Ashoori} systems with inertial dynamics and soft electrostatic edges and are usefully reviewed in reference\cite{Mikhailov}. It is of fundamental interest to observe them in graphene which has non-inertial, ultra-relativistic type dynamics and a hard edge, both from the point of view of the excitations themselves and of the light they can shed on graphene. They show promise too of forming the basis of chiral plasmonics.

Graphene is a monatomic layer of sp$^2$ bonded carbon atoms which form a two dimensional planar honeycomb array. Its electron dynamics is determined by the linear crossings of the band energies at two degenerate but inequivalent points $K$ and $K^\prime$ on the Brillouin zone edge which set the Fermi level for global charge neutrality \cite{Wallace}. The linear dispersion around these points indicates constant speed (magnitude of velocity) ultra relativistic-like massless dispersion. Neutrality can be broken by capacitive charge transfer to explore a reciprocal space Fermi circle on either side of the crossing points. Although the remarkable through-transport properties of this system have been much studied, particularly in the Quantum Hall Effect (QHE) regime that led to the positive identification of single layer graphene \cite{Novoselov,Kim}, few experiments have as yet explored the low energy collective excitation spectrum. Monochromatic laser source experiments \cite{Koppens,Basov} have probed the wavelength of room temperature zero-field plasmons by observing the interference pattern between a point excitation source and reflection from a line boundary. Broadband far infrared experiments\cite{Ju} on periodic strips of graphene have given independent information on density and wavevector dependence of room temperature zero field plasmon absorptions. Other infrared experiments show the spectral behaviour in magnetic field of an array of dots \cite{Avouris} and the field dependence of the absorption lines for a continuous sheet of graphene in the presence of substrate disorder \cite{Orlita}. The latter two works revealed modes decreasing in frequency with magnetic field in accordance with an evolution towards a low frequency, possibly edge localised mode as observed in conventional 2D electronic systems (2DES) with Newtonian dynamics.

The present work investigates directly how an electrical perturbation applied on the edge of an exfoliated sample of graphene in the QHE regime is transmitted to a local probe placed further along the edge. The electrical perturbation, set up by a fast voltage pulse, creates an EMP wavepacket of edge charge whose time of flight to another edge electrode is recorded. The technique\cite{Ashoori} enables an immediate demonstration of chirality and edge propagation, a measurement of the propagation velocity and an estimate of attenuation. The experiment reveals a branch of low frequency gapless modes which propagate unidirectionally around the edge with exceptionally low loss at a velocity which is the sum of single particle drift and Hall conductance whose sign (product of charge and applied magnetic field direction) determines the chirality. It is a good tool for investigating the nature of the edge and shows promise of opening up a new branch of chiral plasmonics with several foreseeable applications for the manipulation of radio to infrared electromagnetic waves.

First results of this work have been reported in reference\cite{emp} and similar work has been reported very recently on edge mode propagation in larger samples at slower time scales in graphene formed on silicon carbide\cite{Kumada}.

\section{Experiment}

\subsection{Principle}

\begin{figure}[h]
\vspace{2mm} \centerline{\hbox{
\epsfig{figure=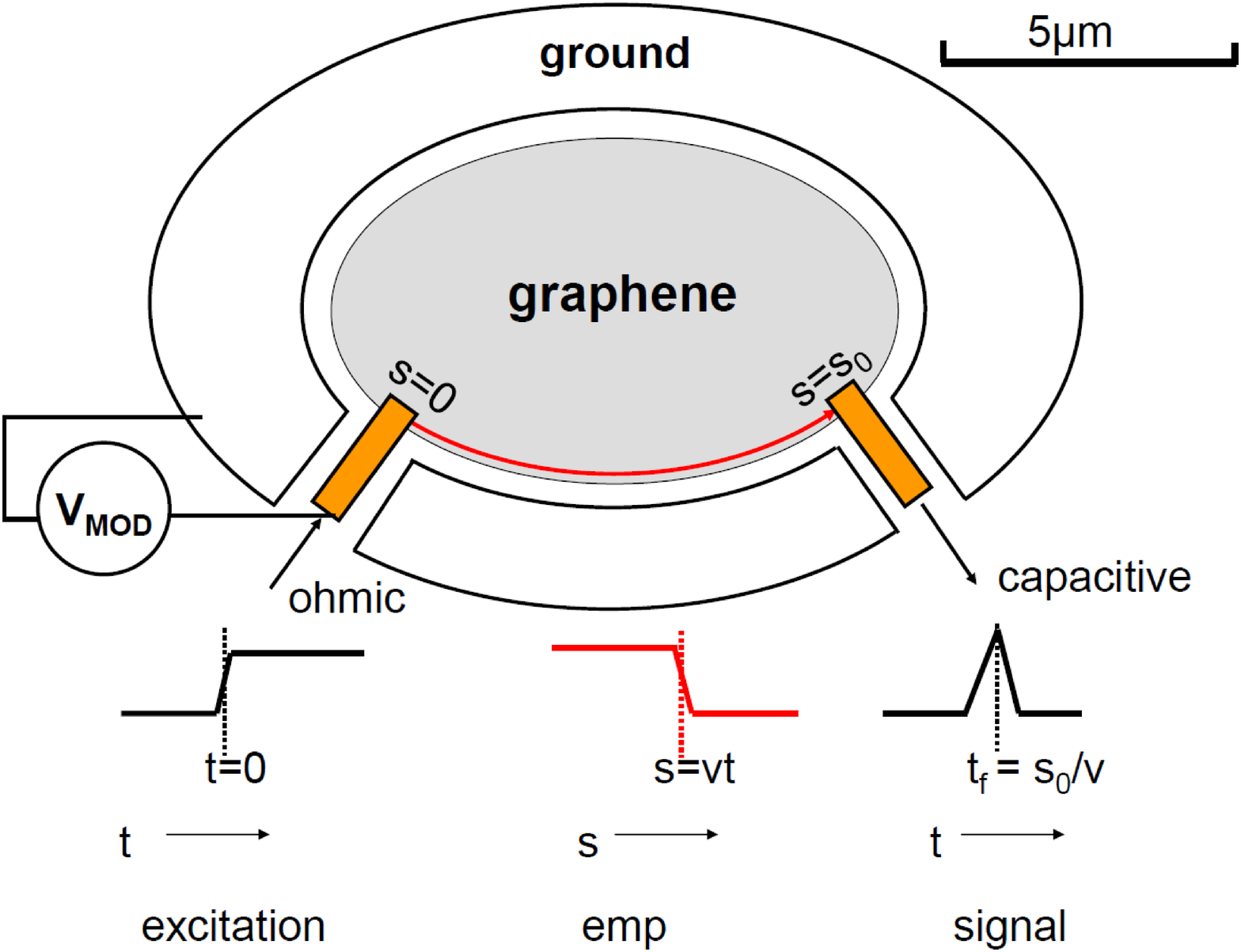,width=80mm}}}
\caption{Schematic diagram of the experimental principle. A fast step excitation voltage pulse is applied through the left hand ohmic contact to create a wavepacket of the excitation, the spatial form of which is illustrated in the middle (red) sketch. Upon arrival at the capacitive receiver contact a time $s_0/v$ later, where $s_0$ is the path length and $v$ the propagation velocity, it is differentiated to give the temporal form sketched on the right. The propagation path is represented by the red arrow along the sample boundary. The gold coloured strips overlapping the edge of the sample by about 2 $\mu$m represent the extremities of tapered coplanar 50 GHz bandwidth transmission lines. Only that part of the signal which is modulated by the sample density variation imposed by $V_{\mathrm{MOD}}$ of the figure is retained in order to eliminate effects of cross talk between transmission lines.}
\label{Fig_Tito}
\end{figure}

The core idea of the experiment is to investigate the propagation of an electrical perturbation between two points on the edge of an exfoliated 
graphene sample. We measured the time of flight of a 7 ps rise time voltage pulse applied to a $\sim 2\times2$ $\mu$m$^2$ electrode on the sample edge to a similar electrode placed about $1/3$ the way around the $\sim 40$ $\mu$m perimeter as indicated in Figure \ref{Fig_Tito}. Although considerably faster and on a smaller sample, the technique is similar to the pulsed EMP experiment on conventional 2DES \cite{Ashoori}. The spatial and temporal definition of the pulse sets up a wavepacket superposition of charge density which propagates out from the source and can be detected by the remote detector as the density waves arrive. The arrival time, spread and amplitude of the signal give information on propagation path, dispersion and scattering of the excitations. We 
interpret the results of the experiment in terms of edge propagating density waves in the Coulomb interacting plasma of excess charge.

\subsection{Fabrication}

\begin{figure}[h]
\vspace{2mm} \centerline{\hbox{
\epsfig{figure=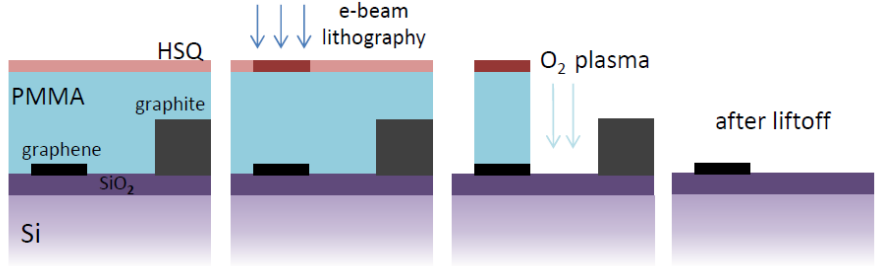,width=120mm}}}
\caption{Removal of graphite from the sample. Panels from left to right: a bilayer of PMMA and HSQ is deposited, e-beam lithography on the positive resist HSQ defines the etching mask, oxygen plasma etching of graphite, liftoff.}
\label{F_Fab}
\end{figure}

\begin{figure}[hh]
\vspace{2mm} \centerline{\hbox{
\epsfig{figure=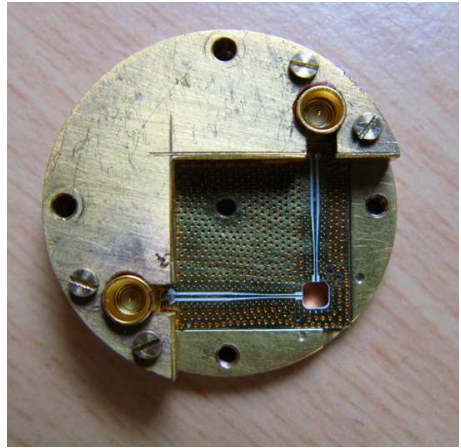,width=60mm}}}
\caption{Sample holder showing coplanar waveguide with the 50GHz bandpass designed for the experiment. The flip chip with graphene is placed over the hole in the bottom right hand corner.}
\label{F_CPW}
\end{figure}

We fabricated the sample by mechanically exfoliating graphene from natural graphite onto a thermally grown 290 nm silicon-oxide surface of a high resistivity Si wafer (8 k$\Omega$-cm at 300 K). The high resistivity  avoids microwave loss and the oxide film facilitates visual identification of single layer graphene\cite{Novoselov2005}. Identification and sample quality were confirmed by micro-Raman spectroscopy which indicated low disorder by the relative areas of the 2700 cm$^{-1}$ and 1600 cm$^{-1}$ Raman peaks\cite{Ferrari}. Spuriously deposited graphite, a possible source of unwanted signal and electrical short circuiting, was removed by oxygen plasma etching using a bilayer of PMMA (PolyMethylMethAcrylate) and e-beam polymerised HSQ (Hydrogen SilsesQuioxane) as an etching mask (see Fig. \ref{F_Fab}). After liftoff only graphene remained on the wafer onto which were then patterned broadband (50 GHz) tapered Ti/Au coplanar waveguides (CPW) by standard e-beam lithography. The extremities of the CPW centre conductors contact the sample edge over a $2\times2$ $\mu$m$^2$ area, one resistively and the other capacitively. A sample photo and further fabrication details are given elsewhere\cite{emp}. The resulting $3\times3$ mm$^2$ sample ``flip-chip" is mounted face down on a sample holder (see Figure \ref{F_CPW}) on which the CPWs are continued to mini-SMP strip-to-coaxial microwave connectors to join the vertical coaxial transmission lines which exit the cryostat insert. Elasticity of the flip-chip to sample holder contacts was carefully engineered to ensure flatness and good electrical connection across the entire CPW (left, right and centre) when pressed into place. Soldering was avoided entirely because the subsequent annealing of the graphene at 150$^\circ $C for several hours would have risked melting the solder and contaminating the sample.

The sample holder, shown in Figure \ref{F_CPW}, was designed using CST Microwave Studio (3D Maxwell equation solver) and fabricated on a high dielectric constant ceramic wafer Rogers TMM10 with copper ground and top planes joined by a set of densely spaced vertical interconnection via holes and thin enough not to support transverse standing modes over the 50 GHz frequency band. 
The assembly was mounted in vacuum in a cryogenic insert which fits into the liquid helium filled bore of a 19 T magnet at 2.2 K.

\subsection{Experimental Procedure}
We conducted time domain measurements using a Tektronix DSA8200 digital serial analyzer and sampling oscilloscope with time-domain reflectometry (TDR) 80E10 (50GHz bandwidth) and 80E08 (30GHz bandwidth) modules in transmission mode. A 7ps (80E10)/11ps (80E08) rise time voltage step was injected into one side of the circuit and the response voltage measured at the other as a function of time. A low frequency (1Hz) square wave voltage is added at the ohmic input via a bias tee in order to modulate the density on the edge with the aim of separating out the density sensistive graphene signal from crosstalk. For each excursion of the modulation the signal is averaged over a maximum of pulse injections ($\sim 10^3$) and the resulting picosecond files subtracted  to demodulate and thereby retain only that part which is sensitive to variation in edge density of the sample. The procedure for extracting arrival times from the demodulated signal is described below.

\subsection{Results}

The absolute signal arrival time includes the travel time through the transmission lines and therefore it is necessary to determine the zero time, i.e. the moment the excitation signal reaches the sample, in order to extract the time of flight. We set the zero by the arrival time for zero field, supposing it to originate in the surface propagation of the fast surface plasmon mode, as there is nothing to impose chirality and the surface mode propagation time is of the order of 1ps, which is comparable with our time resolution.

\begin{figure}[hh]
\vspace{2mm} \centerline{\hbox{
\epsfig{figure=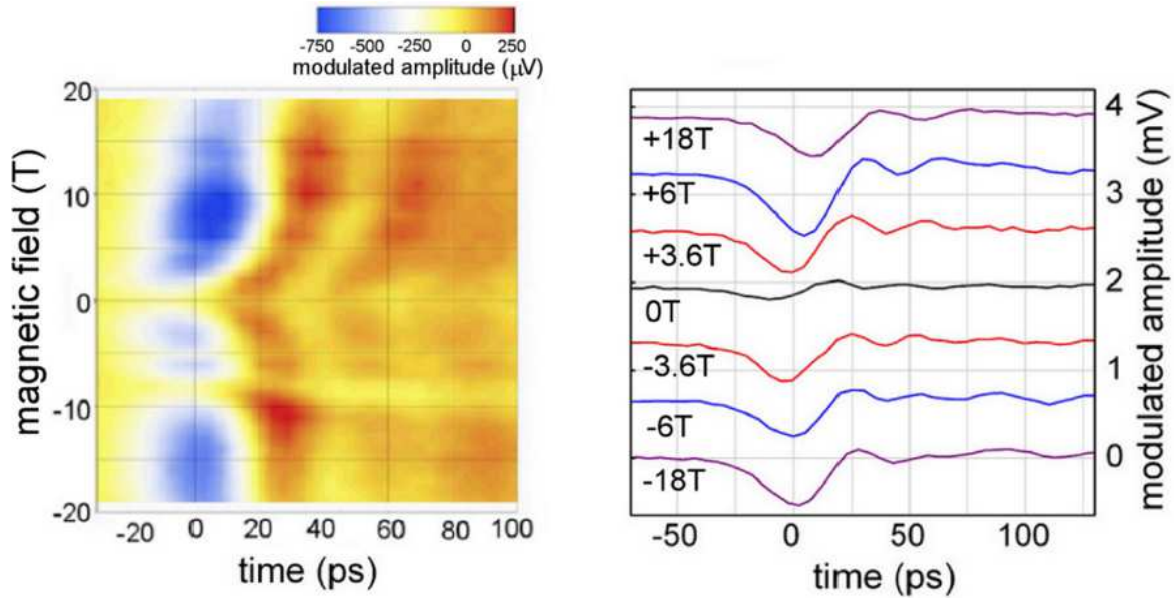,width=160mm}}}
\caption{Transmitted voltage as function of time and magnetic field. Left: colour plot of signal amplitude. Right: traces for several field values, corresponding to the integer filling factors $\nu = 2,6,10,\infty$ \cite{emp}}.
\label{F_data}
\end{figure}

\begin{figure}[hh]
\vspace{2mm} \centerline{\hbox{
\epsfig{figure=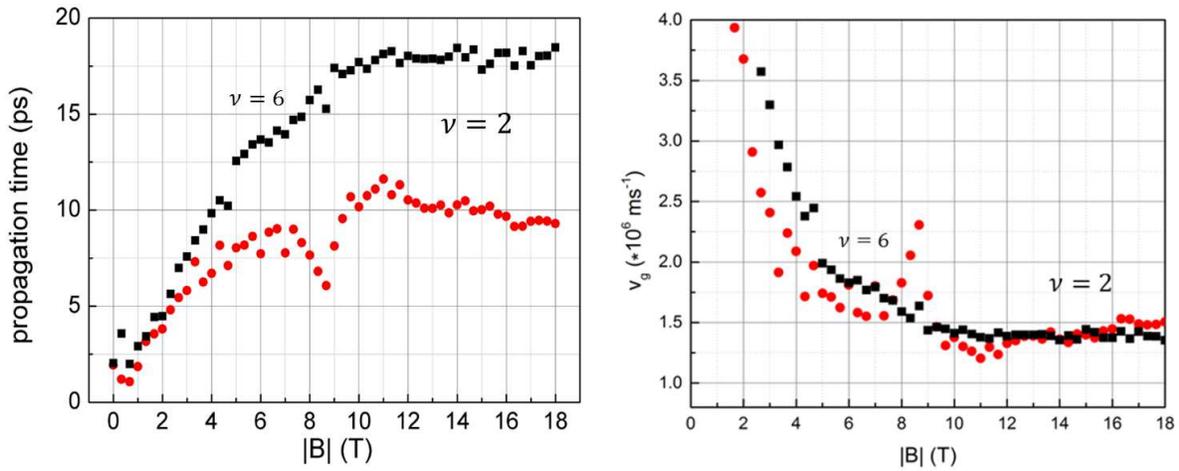,width=160mm}}}
\caption{Left: Propagation time as function of field. Right: Propagation velocity as function of field\cite{emp}.}
\label{F_main}
\end{figure}

We employ the fitting procedure described below to extract the signal propagation times for the longer and shorter paths as a function of field and we find that they scale as the path lengths (see Figures \ref{F_data},\ref{F_main}). Taken together with the absence of a simultaneous counter-propagating signal, the scaling shows clearly the unique chirality of the modes.

The propagation time is a non linear function of field and exhibits a plateau-like structure (see Figure \ref{F_main}) which we attribute to the onset of the quantum Hall effect. We are able to identify the $\nu=2$ and $\nu=6$ plateaus and although we were unable to make DC transport measurements on this sample to confirm the plateau sequence, we have done so on similarly fabricated samples which all show the characteristic onset of a very wide $\nu=2$ plateau characteristic of graphene.
Since the $\nu=6$ plateau is narrow and centered on 6T we infer the centre of the $\nu=2$ plateau to be at 18 T and tentatively pick out the $\nu=10$ to be at 3.6 T although clear identification of this latter plateau is lacking. A seemingly anomalous feature is the dip at 8 T in the propagation times for the shorter path only. We were unable to ascertain the origin of this dip.

\begin{figure}[hh]
\vspace{2mm} \centerline{\hbox{
\epsfig{figure=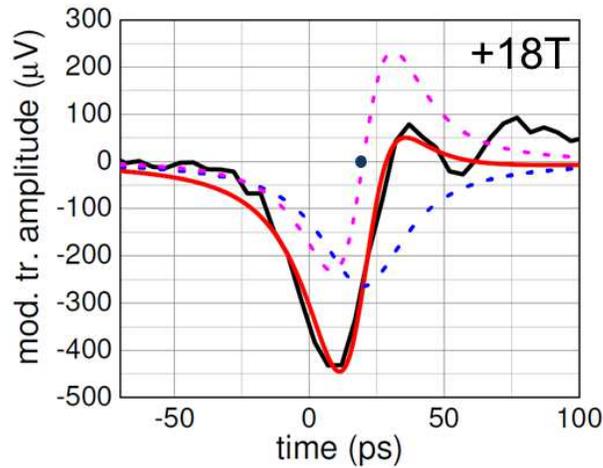,width=80mm}}}
\caption{Example of functional fit for extracting arrival time. The experimental signal in black is fitted to a weighted sum of first (dashed blue) and second (dashed pink) derivatives of a Fermi function representing modulation of attenuation and velocity respectively. The blue filled dot indicates the arrival time. See text.}
\label{F_signal_shape}
\end{figure}

Arrival times are extracted from the demodulated signal by a functional fitting procedure.  A typical measured modulated voltage as function of time  is given in Figure \ref{F_signal_shape} in black. The excitation step voltage is not an ideal Heaviside function but has exponentially rounded corners which are well represented by a Fermi function $f(t)=(\exp\alpha(t-t_0)+1)^{-1}$, where $\alpha$ is a parameter set by the rise time and $t_0$ the position of the step. We suppose this to set up an EMP wavepacket of the same shape which propagates as sketched in red in Figure \ref{Fig_Tito}, later times appearing closer to the injection point. When this wavepacket arrives at the receiver electrode it is transformed into a time domain signal which is differentiated by the RC of the capacitive contact. The slow uniform modulation of edge density has a twofold effect: it can modulate the amplitude of the wavepacket, which does not change its arrival time, and it can modulate its propagation velocity which changes its arrival time, but not its amplitude. Taking account of the additional RC differentiation, the attenuation action gives rise to a symmetric first derivative and the velocity action to an antisymmetric second derivative in the demodulated signal (supposing the action of each to be linear in modulation voltage). The relative contributions of the two effects can differ in different conditions, but the arrival time is common to both. We therefore fit each demodulated signal individually to a weighted sum of first and second derivatives of a Fermi function with a common arrival time as shown in Figure \ref{F_signal_shape} where an experimental signal is shown in black and the two derivative components in dashed blue and pink, their sum being the fit in red. The arrival time is represented by the blue dot. Modulation acts differently on and off a plateau since the relative contributions of attenuation and propagation velocity vary, resulting in a change of shape with field.

The relative amplitude of the signal along the longer and shorter paths informs us on attenuation over the difference in path lengths and we are able to estimate roughly an attenuation length of 70 $\pm$ 30 $\mu$m corresponding to a relaxation time of 50 $\pm$ 20 ps, three orders of magnitude longer than the Drude relaxation time relevant for the damping of surface plasmons as deduced from the DC mobility $\mu = e\tau /m^* \rightarrow \frac{ev_F}{\hbar k_F} \tau$ \cite{Das Sarma} measured on similar samples (see below for origin of effective inertial mass $m^*$).


\subsection{Summary}

The experiment has shown that:

\noindent (i) The perturbation is transmitted from one contact to the other with a time delay that is proportional to the perimeter path length, the shorter or longer path being chosen by the sign of the applied perpendicular magnetic field. There is no evidence for simultaneous propagation along both. This establishes chirality and edge propagation.

\noindent (ii) The time of flight is a function of magnetic field and shows quantisation on QHE plateaus. This identifies the excitation to be associated with the QHE.

\noindent (iii) The shorter path length gives a larger signal amplitude. This enables us to estimate the attenuation length of the wave by comparing the relative signal amplitudes for left and right paths.

The ensemble of data features is well described in terms of edge magnetoplasmons.

\section{Plasmons}

\subsection{Surface plasmons}

It is helpful to situate the edge magnetoplasmons in the more general context of plasmons. Plasmons are propagating charge density waves whose restoring force arises from Coulomb interaction. The Fourier transform $E^{(D)}_k \sim k^{(2-d)} n^{(D)}_k$ in dimension $D=d$ of the Coulomb field from charge density variation $n_k ^{(D)}\cos \textbf{k.r}$ is the root of the  dimension sensitive frequency-wavevector dispersion: finite dispersionless frequency in 3-, square root in 2- and linear in 1-dimension. Graphene has two-dimensional ultra-relativistic type dynamics according to which a force cannot change the magnitude $v_F$ but only the orientation $\theta$ of the velocity which is always aligned with the particle momentum. The momentum $\textbf{p}$ is tilted by $\delta \theta \sim Fdt/p$ by the action of a force $\textbf{F}$ for a time $dt$ leading to a change in velocity $dv \sim v_F \delta \theta \sim (v_F /p)Fdt$ which, somewhat paradoxically, behaves like an effective inertial mass $m^\ast \sim p_{F}/v_{F}$. Although this does not change the wavevector dependence, $\omega \sim k^{1/2}$, the appearance of the density dependent Fermi momentum $k_{F}=\sqrt{\pi n_s}$ does alter the density dependence from $\omega \sim \sqrt{n_s k}$ for inertial 2DES dynamics to $\omega \sim \sqrt{n_s ^{1/2}k}$ and in practice scales the dispersion to much higher frequency. A fuller, quantum description can be found in reference\cite{Das Sarma2}.

We attribute the arrival time at zero field in Figure \ref{F_main} to surface plasmon propagation whose group velocity diverges as $k^{-1/2}$.

\subsection {Edge plasmons}
In perpendicular magnetic field the dynamics is dominated by mass independent Lorentz force drift, as in conventional 2D electron systems (2DES), because fundamentally it only depends on the Lorentz transformation to the electron velocity frame. We can therefore transfer to graphene much of the insight gained from the study of inertial 2DES. As observed in those systems, upon applying a magnetic field perpendicular to the graphene the longitudinal plasmon modes split into frequency increasing and frequency decreasing branches. While for the former the wave amplitude remains spread over the whole interior of the sample, for the latter it progressively confines itself to the sample boundary. This is the creation of the edge magneto-plasmon (EMP) mode with which we identify the observed propagation of edge charge.

This low frequency gapless ($\omega \rightarrow 0$ as $q\rightarrow 0$) mode is a quasi one dimensional charge density wave propagating around the edge of the electron pool. In a classical description\cite{Dahm, Glattli,Mikhailov, Volkov88} an EMP wave propagates by depositing charge on the edge by virtue of internal drift currents set up by the field of its own periodic edge charge distribution. The particle motion associated with the wave is Lorentz drift in the superposition of the single particle constraining edge potential and the 
electrostatic interaction potential of the edge charge density variation. The velocity $v_D$ of the drift of electrons along the edge potential determines a reference frame for the Hall conduction propagation, adding to its propagation velocity when seen from the laboratory. The Hall conductance contribution always dominates the propagation velocity in the usual semiconductor heterojunction 2DES except in the fractional QHE regime\cite{Wassermeier} but not, as we shall see, in graphene. The currents are confined to a depth given by the shorter of the screening length or the wavelength. They propagate unidirectionally along the edge with chirality determined by the Lorentz drift, imposing propagation in a direction set by the sign of the Hall conductance $\sigma_{xy}=\frac {n_s ec}{B}\equiv \nu \frac{e^2}{h}$ if the filling factor is regarded as carrying the sign of the product of the signs of excess charge and magnetic field. They propagate at group velocity
\begin{equation}
v_{g}=\frac{d\omega}{dq}=\frac{2\,\sigma_{xy}}{\epsilon_{\rm eff}}\left( \ln\frac{2}{\left\vert
\,q\right\vert w}+C-1\right)+v_{D},
\label{eq2}
\end{equation}
\noindent where $\epsilon_{\rm eff}$ is the effective dielectric substrate screening constant, $q$ the wave vector, $w$ a lower cut off length associated with effective penetration of the distribution of charge accumulated at the edge or by the screening length set by the mobility of the compressible parts of the 2D gas. $C$ is a density profile sensitive constant of order 1\cite{Volkov88}. 

We associate the higher field ($B>2.5$ T) part of Figure \ref{F_main} with these excitations.

\subsection{Electronic structure of the edge}

A more detailed analysis of the EMP propagation velocity calls for a model of the edge. It will  situate the relative importance of the single particle and interaction (Hall effect) contributions and make more explicit the quantum description of EMPs.

Inspired by Halperin's early work\cite{Halperin} we think of the graphene edge, i.e. the border of the carbon network, as a wall beyond which the electron wavefunction must vanish. By virtue of being squeezed between the sample edge and the position set by the wavevector parallel to the edge (in the Landau gauge), it acquires extra quantum confinement energy. The electronic edge is then not the sample edge but rather the position at which the energy of the lowest Landau level crosses the Fermi level. The characteristic energy for the crossing, $\hbar v_F/ \ell_B$, is reached over a magnetic length $\ell_B = \sqrt{\hbar c/eB}$ (the width of a wavefunction) from which it is straightforward to estimate $v_D=cE_c/B\sim v_F$, the electric confinement field $E_c$ representing the slope of the Landau level energy on approaching the edge. The interaction (QHE) contribution to the velocity at the lowest integral filling factor $\nu = 2$ is proportional to $2e^2 /h\epsilon_{\rm eff}$ and $\approx 7 \times 10^7$ cm s$^{-1}$, comparable to the Fermi velocity $v_F \approx 10^8$ cm s$^{-1}$. It is interesting to note that, for conventional 2DES too, $v_{D} \sim v_{F}/\nu^{1/2}$, but there $v_F$ is set by the average electron spacing rather than the band crossing slope which is related to the carbon-carbon spacing of the  graphene lattice.

Thus far we have implicitly assumed that the value for the electronic density at the edge is that associated with the stationary QHE on the grounds that the edge is also where the transport current flows and the Hall charge accumulates. To link it to the value in the interior we have to take into account the charge distribution across the sample which is set principally by electrostatics modified somewhat by the Landau level structure \cite{SCG}. In our configuration there is no screening back gate, only a surrounding ground plate a few micrometres distant by which we electrostatically modulate the charge density. If we were to think of the sample as a conducting disc of radius $R$ without nearby screening electrodes, the electrostatic solution for excess charge density as a function of the radial coordinate $r$ would be $n(r)=\frac{n_0}{(1-(r/R)^2)^{1/2}}$ which diverges at the edge as $(R-r)^{-1/2}$ as it must on approach to the edge of any unscreened 2D conductor, at least over a distance comparable to that to the nearest screening electrode. The implicit assumption in this electrostatic result is that the charges are held in  by the edge of the electrode. This is to be contrasted with a depletion edge commonly met in conventional 2DES where an outside electrode imposes a potential which would call for a charge deficit at the edge in order for the sample to remain everywhere a conductor. Unlike a simple metal, mobile excess charge of a semiconductor with a gapped electronic band structure, or mobile charge on an insulating substrate, can be pushed away from the edge leaving a charge depleted insulating strip until the potential reaches the point where charges from the band on the other side of the gap can furnish charge of opposite sign (inversion). The general form of the charge density on approach to a depletion edge, $n(r)\propto (R-r)^{1/2}$ is very different. Graphene, however, has no gap, only a crossing (Dirac) point at which it remains conducting. Evolution from excess to deficit (hole) charge is continuous as in a metal and there is always a density divergence on the edge (except for the special point where the edge coincides with charge neutrality). In magnetic field, the electrostatic solution minimising the Coulomb energy still globally dominates, but the Landau level structure does modify the electrostatic density distribution locally on either side of areas of complete Landau level filling where the addition of a particle requires a Landau level spacing energy. This gap incompressibility leaves constant density, filled Landau level, narrow strips within which no screening occurs and the potential varies by the Landau energy gap. They separate wider, unfilled Landau level, constant potential varying density bands across the sample, but the smoothed average of the density follows closely the electrostatic distribution \cite{SCG,SE}. Quantum Hall plateaus result when the Fermi level lies between Landau levels in the sample interior and the filled levels emerge from the Fermi sea at the edge. The EMP modes are carried along these emergence lines as QHE excitations.

\subsection{Quantum description of EMP and drift velocity}

The quantum description of the EMP wave can now be understood as a propagating variation in local chemical potential which, according to the QHE, is accompanied by a periodic Hall current. The spatial variation of the current redistributes the charge along the edge by shrinking or expanding the electron pool \cite{Wassermeier}. The local chemical potential is determined by the combination of the electrostatic interaction and position of the electronic edge in the  the single particle quantum confinement energy of the Landau level. The result for the velocity in the simple case of a single conduction channel $i$ is the same as in the semi-classical picture outlined above, $v_D ^{(i)}=\frac{c}{eB}\frac{\partial W_{LL} ^{(i)}}{\partial r}|_\mu $ (in terms of the quantum capacitance per unit length $C^{Qi}=\frac{\partial N^i e/L}{\partial \mu /e}=e^2\frac{B}{\Phi _0}\frac{dr^i}{d\mu ^i}$, $v_D ^{(i)}= \frac{\sigma _0}{C^{Qi}}$ for each channel $i$, $\sigma_0 =e^2/h$ being the conductance quantum). $W_{LL} ^{(i)}=W_{LL} ^{(i)}(r)$ is the energy of the corresponding Landau level and the derivative is evaluated at the equilibrium Fermi level $\mu =\mu_F$. Charge spread is expected to be of order of the magnetic length $\ell_B$.
When $\nu \geq 1$, multiple Landau levels emerge through the Fermi level and the question arises of how to take into account differing drift velocities of multiple channels. If the differences of drift velocities between channels are small compared with the EMP velocity, the drift velocity for the EMP mode associated with the $\nu$ interacting channels is a simple average $\langle v_{D}\rangle=\nu^{-1}\sum \limits _{i=1}\limits ^{\nu} v_{D}^{(i)}$ of the drift velocities at the points where each Landau level emerges through the equilibrium Fermi level (in the language of quantum capacitance $\langle v_{D}\rangle=\sigma _0 \nu^{-1}\sum \limits _{i=1}\limits ^{\nu}\frac{1}{C^{Qi}}$). The group velocity would then become
\begin{equation}
v_{g}=\frac{d\omega}{dq}=\frac{2\,\sigma_{xy}}{\epsilon_{\rm eff}}\left( \ln\frac{2}{\left\vert
\,q\right\vert w}-\gamma-1\right)+\langle v_{D}\rangle
\label{eq3}
\end{equation}
if all the levels are supposed to emerge at the same position. The electrostatic potential has here been assimilated to that from a circular wire of diameter $2w$ and $\gamma\approx0.577...$ is the Euler constant. The degrees of liberty introduced by multiple Landau levels allow for higher order modes, analogously to the classical system\cite{Aleiner-Glazman,Nazin-Shikin}. In the quantum picture they are described by different weights of chemical potential variation over the  multiple Landau level channels. The EMP mode corresponds to in phase charge oscillations over all the channels together and constitutes the fastest mode. The higher, multipolar modes propagate more slowly, approximately at drift velocity little influenced by the EMP mode. They are however not expected to be excited by our electrodes which are considerably larger than the magnetic length. The characteristic length in the log of Equation 2 results from the electrostatic potential of a strip of width $w$ with periodic charge distribution along its length. $w \sim \ell_B$ in the quantum picture. 
Equation 2 supposes all the Landau levels to emerge from the Fermi sea at the same location and to have the same width, but in reality they emerge in groups corresponding to the Landau level indices. The effect of taking into account the separation between different groups is to reduce the electrostatic contribution due to the separation. For our situation, the reduction is estimated to be of order 10\% for 
$\nu$=6 and 20\% for $\nu$=10\cite{PWG}.

\subsection{Calculation of drift velocities in graphene for a hard wall}

\begin{figure}[hhh]
\vspace{2mm} \centerline{\hbox{
\epsfig{figure=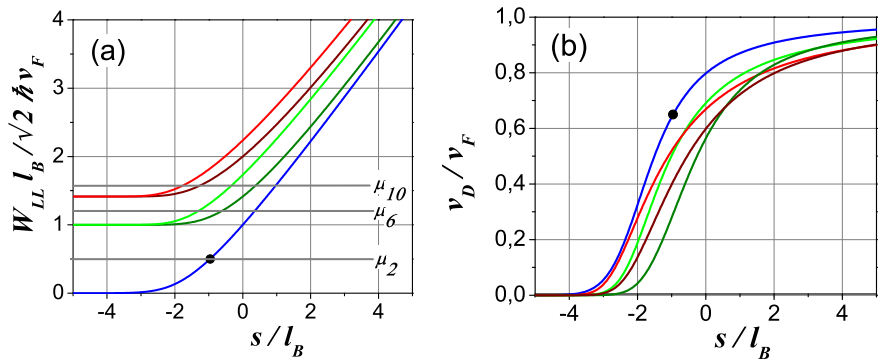,width=140mm}}}
\caption{(a) Left panel: Landau level energies for graphene in the presence of a hard wall potential as a function of distance to the edge. (b) Right panel: slope of energies in panel (a) from which the drift velocity for each channel is evaluated by transposing the crossing points between the Landau level energies in (a) with the appropriate Fermi level $\mu _\nu$. The example of $\nu=2$, $N=0$ is illustrated by the two black dots. The velocities are recorded in Table \ref{table:velocities}.}
\label{F_LL_VD}
\end{figure}

Berry and Mondragon solved the Dirac equation with perpendicular magnetic field considering hard wall confinement of the sample edge which enters into the equation as an infinite mass term, rather than an infinite electrostatic barrier\cite{Berry}. They also formulated the appropriate boundary conditions 
which for graphene are different for a zigzag or an armchair edge\cite{Beenakker_edge}. We consider a zigzag edge which we believe to be more appropriate to a random edge obtained by oxygen plasma etching. Following this approach and along the lines adopted in other papers \cite{Fertig&Brey,Abanin,thatRussianone} we calculate numerically the Landau level energies $W_{LL}$ for graphene for a hard wall and a zigzag edge as a function of distance to the edge as shown on the left hand panel of Figure \ref{F_LL_VD}. The splitting of levels at the edge corresponds to lifting of valley degeneracy. The right hand panel shows the spatial derivative of the energies from which the drift velocities $v_D^{(i)}=\frac{1}{\hbar}\frac{dW_{LL}^{(i)}}{dk}=\frac{c}{eB}\frac{\partial W_{LL} ^{(i)}}{\partial r}$ are deduced by 
transposing the positions at which the energy levels cross the Fermi level in Figure \ref{F_LL_VD}(a) to the derivative curves in Figure \ref{F_LL_VD}(b). The Fermi level $\mu$ for a given filling factor $\nu$ is taken to be midway between the Landau levels of indices $N=(\nu - 2)/4$ and $N+1$ far from the edge as appropriate to a QHE. The drift velocities for the  valley split branches are evaluated individually for each quantum Hall plateau. 
The results for the positions and individual drift velocities for $\nu=2,6,10$ as well as their averages are recorded in Table \ref{table:velocities} together with the measured group velocities and the Hall effect coefficient $v_0 = \frac{\nu\sigma_0}{\varepsilon _{\rm eff}}$ multiplying the $(\ln ({2/qw}) - \gamma - 1)$ term of Equation 2. The Landau levels in play for each filling factor $\nu$ are labeled by their indices $N$ and the broken valley degeneracy is shown explicitly whereas spin degeneracy is not; $\Delta_{10}$ is the energy splitting between the $N=0$ and $N=1$ Landau levels and $v_F \approx 10^8$ cm s$^{-1}$. 

\begin{table}
\caption{\label{velocities}Measured group velocity $v_g$ and calculated Hall velocity coefficients $v_0$, Landau level positions $s$ and drift velocities $v_D$ at Fermi level $\mu =\mu_{\nu}$.}
\centering
\begin{tabular*}{1\textwidth}{@{\extracolsep{\fill}}c c c c c c c c}

\hline\hline
$\nu$&$N$&$\mu/\Delta_{01}$&$v_g$ ($10^8$ cm s$^{-1}$)&$v_0$ ($10^8$ cm s$^{-1}$)&$s/\ell_B$&$v_D/v_F$&$\langle v_D/v_F \rangle$\\ [.75ex]
\hline
&&&&&&&\\
2&0&0.500&$1.4\pm0.2$&0.21&-0.97&0.65&0.65\\
&&&&&&&\\
6&0&1.207&$1.8\pm0.2$&0.64&+0.34&0.83&\\
&1&&&&-0.83&0.40&0.55\\
&&&&&-1.29&0.42&\\
&&&&&&&\\
10&0&1.573&$2.5\pm0.3$&1.07&+0.97&0.87&\\
&1&&&&+0.35&0.64&\\
&&&&&-0.33&0.64&0.56\\
&2&&&&-1.27&0.33&\\
&&&&&-1.77&0.34&\\
\hline
\end{tabular*}
\label{table:velocities}
\end{table}

Table \ref{table:parameters} illustrates the different steps in our analysis of the data. The initial step is to derive the log term of Equation 2 by fitting $v_g = \beta v_0 + \langle v_D \rangle$ to the measured velocities $v_g$ at each filling factor upon taking $\langle v_D \rangle$ from the above calculation. Equation 2 identifies $\beta$ with $(\ln (2/qw) -\gamma-1)$ enabling us to evaluate $qw$. $q$ is deduced from the pulse rise time which we assimilate to one side of a Gaussian in time which we convert to a spatial Gaussian of half width $\ell=v_g \tau$, the Fourier transform of which gives a Gaussian distribution of half width $q=\ell ^{-1}$. Given $q$ we go on to evaluate the effective width $w$. In the last column of the table we compare $w$ with the magnetic length $\ell_B = \sqrt{\hbar c/eB}$ at the field of the experiment($\ell _B \sim 25.6$ nm for 1 tesla). Because the exponential is strongly non linear over the range of the experimental error limits, we have given the asymmetric limits for $w$ on either side of the median value corresponding to the extremes of the experimental error estimations. 
The argument of the log term, which is the ratio of the EMP wavelength$/2\pi$ to the unspecified effective width $w$, is expected to be $\gg1$. As is clearly seen in the table, the slow log dependence in that region has the consequence that the relative precision in $qw$ is much reduced over the precision in the measured group velocity (limited primarily by the temporal resolution of our finite bandwidth setup). From Table \ref{table:parameters} we see a window of $200<w<700$ nm, some 30-50 times the magnetic length $\ell_B$. Even though it scales roughly like $\ell_B$ we feel that because the factor is so large it is probably unconnected.

\begin{figure}[hhh]
\vspace{2mm} \centerline{\hbox{
\epsfig{figure=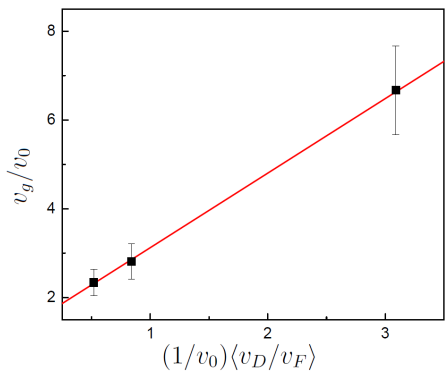,width=60mm}}}
\caption{Measured EMP group velocity as function of the calculated drift velocity for the three filling factors $\nu=2,6,10$. Both quantities are scaled by the filling factor dependent velocity $v_0(\nu)$. The slope of this curve is the Fermi velocity $v_F$ and the y-axis intercept is related to the width of the electrostatic edge $w$, see text. }
\label{F_8}
\end{figure}

An alternative method to analyze the data was to keep the product of the edge width and wave vector $q w$ and the Fermi velocity $v_F$ as two fitting parameters. By dividing Eq. (2) by $v_0$ we obtain $v_g/v_0=\kappa+(1/v_0)\langle v_D/v_F \rangle v_F$, where $\kappa=\ln (2/qw)-\gamma-1$. We measure $v_g$, calculate $\langle v_D/v_F \rangle$ and keep $\kappa (q w)$ and $v_F$ as fitting parameters. In this case the fit turns out to be rather good in terms of linearity (see Figure \ref{F_8}), but overestimates the Fermi velocity to $1.7*10^8$ cm s$^{-1}$ and we find $q w=0.1$. We have estimated $q=14 \mu m \:(\nu=2), 18 \mu m \:(\nu=6), 25 \mu m \:(\nu=10)$ from the spatial width of the measurement pulse rise time as mentioned previously, and that yields $w$ of the order of 1 $\mu m$.

Large $w$ represents slow propagation, slower than expected by a factor of $\sim \ln ({w/ \ell _B}) \sim 3.5$. Although this may be good news for edge plasmonics (see next section), it is unsatisfactory for our understanding. 
The reduced Coulomb interaction between channels mentioned earlier would only reduce $v_0$ by about 10\% for $\nu$=6 and 20\% for $\nu$=10.
A less fundamental, though technically important possibility is that the edge formed by the oxygen plasma etch meanders across $w\sim 500$ nm on length scales $\sim q^{-1}$ over which the electrostatic potential averages. That possibility could be checked using an imaging technique like atomic force microscopy or alternatively by doing EMP experiments on well defined regular edges. 
As concerns the importance of the drift term, we remark that if it 
were absent altogether 
the argument of the log term would have to change by three orders of magnitude in order to account for the measured change of the group velocity between filling factors 2 and 10, posing an even greater challenge to our understanding.

\begin{table}
\caption{\label{paramaters}Parameters extracted from experimental results}
\centering
\begin{tabular*}{1\textwidth}{@{\extracolsep{\fill}}c  c  c  c  c}

\hline\hline
$\nu$&$\beta$&$1/qw$&$w$(nm)&$w/\ell_B$\\ [0.5ex]
\hline
2&$3.6\pm0.95$&$222>85>33$&$63<160<400$&$10<25<70$\\
6&$1.97\pm0.35$&$25>17>12$&$300<1000<1500$&$30<100<150$\\
10&$1.87\pm0.35$&$22>16>11$&$1000<1500<2000$&$70<100<150$\\[1ex]
\hline
\end{tabular*}
\label{table:parameters}
\end{table}

\subsection{Chirality and scattering}

The Lorentz force dynamics imposes chirality. The low frequency, edge magnetoplasmon branch is confined to the edge and travels to the left or right 
according to the sign of the Hall conductance set by the product of the signs of perpendicular field (up or down) and excess charge (electrons or holes). The other, high frequency, branch is delocalised over the sample, has opposite chirality and is gapped by the cyclotron frequency. This raises questions on when chirality sets in and its effect on the attenuation of the wave. Guidance can be sought from the classical analysis of a circular sample \cite{Glattli} where opposite chiralities are labelled by positive and negative azimuthal quantum numbers $\pm {m}$ whose frequencies are degenerate at zero field where they combine to form a longitudinally propagating non-chiral surface plasmon mode. One of these states evolves with magnetic field to localise on the edge while the other remains delocalised over the surface and participates in the gapped magnetoplasmon branch. Such a pair of states can be mixed by an electrical potential which breaks the continuous rotational symmetry with a $2m$-polar moment produced by a random set of impurity scattering potentials. We are therefore led to expect that chirality and confinement to the edge set in properly only once the magnetic splitting becomes large in comparison with the random potential created by the impurities. Once it has established its unidirectionality and has become confined to the edge it has no counter-propagating wave into which it can elastically scatter since the counter-propagating interior modes all have higher frequencies as long as $\omega _{plasmon}<\omega_{cyclotron}$. The EMP modes can then no longer be attenuated by energy conserving back scattering. In contrast, if there were a gapless branch propagating in the backwards direction strong elastic backscattering off random impurities or roughness could result. As it is, elastic backscattering can only occur into the opposite edge, but even that is considerably reduced by virtue of confinement of the EMP to the edge and the separation between edges.



\section{Chiral plasmonics}

The chirality, the ease of electrical control and the low attenuation of EMP in graphene point to interesting applications in the rising field of plasmonics. Three port miniaturised microwave circulators become possible in the GHz to THz range and particularly flexible by virtue of the electrical tuning possibility: the propagation velocity can be varied linearly with gate voltage and even have its chirality reversed in a few picoseconds by the flip of a switch. This feature also confers the possibility of using the circulator as a microwave switch between a left or right port and a two port device could be made into a  voltage variable phase shifter. These applications would require a static magnetic field of a few tesla, but only over the extent of the sample which can be restricted to a few tens of microns. Inasmuch as the propagation impedance of the wave is the Hall resistance $R_{H}=\nu ^{-1}h/e^2\approx12.5/\nu$ k$\Omega$, however, the problem of impedance matching the device to external electromagnetic transmission lines would have to be overcome. Nonetheless EMP based devices would seem to have certain advantages over zero-field plasmon based devices in the much reduced attenuation and in the linear rather than fourth root gate voltage dependence of propagation velocity, although impedance matching would be somewhat easier for surface plasmons in that it can be tuned by varying the width of the sample rather than the number of Landau channels on the edge.

\section{Conclusion}
The experiments have demonstrated that edge magneto-plasmons exist on the perimeter of graphene samples and that they are described by Hall effect dynamics superimposed on the single particle Lorentz drift perpendicular to the force field created by the quantum confinement energy of the Landau level upon approach to the sample edge. The velocity of the drift is found to be close to the Fermi velocity and of the same order as the Hall conductance contribution to the propagation velocity at the lowest filling factor $\nu=2$. They have been shown to be chiral with a relaxation time three orders of magnitude longer than that of non-chiral, zero magnetic field plasmons. In the quantum Hall effect regime, the propagation velocity is quantized in the same manner as the Hall conductance. Both the single particle drift and the Hall contributions are sensitive to the structure of the edge, a feature of considerable interest for distinguishing zig-zag from armchair edges when these become available in pure, well defined form. On a more applied front, these excitations extend the already promising applications of graphene for plasmonics to chiral plasmonics for use in wide bandwidth, voltage tunable and reversible devices such as circulators and phase shifters.

\section{Acknowledgements}
We are grateful for input from Keyan Bennaceur, Fabien Portier and Patrice Roche. We gratefully acknowledge discussions with Leonid Glazman, Barry Bradlyn, Jack Harris, Sergey Mikhailov and Allan MacDonald. We are indebted for excellent technical help to Patrice Jacques, Claudine Chaleil and Pierre-Fran\c{c}ois Orfila. This work was supported mainly by the ERC Advanced Grant MeQuaNo No. 228273 and also by the RTRA Gamet Grant No. 2010-083T, the ANR PNANO grant ``METROGRAPH" and the ANR white grant ``SUPERGRAPH".

\end{document}